\documentclass[11pt,a4paper]{article}
\pdfoutput=1

\usepackage{amsfonts,setspace}
\usepackage{verbatim}
\usepackage{float}
\usepackage{amssymb}
\usepackage{amsmath}
\usepackage{latexsym}
\usepackage{graphicx,color}
\usepackage{caption}
\usepackage{graphicx}
\graphicspath{ {images/} }
\usepackage{subcaption}
\usepackage{mathrsfs}
\usepackage[colorlinks=true, allcolors=blue]{hyperref}

\newcommand{\zm}{z_{-}}
\newcommand{\zp}{z_{+}}
\newcommand{\z}{z_{\pm}}

\usepackage[margin=0.85in]{geometry}
\hypersetup{colorlinks=false, linkcolor=blue, citecolor=red}
\newcommand{\vs}[1]{\vspace{#1 mm}}
\begin{document}

	\begin{flushright}
	\end{flushright}
	
\begin{center}
		{\huge{\bf Revisiting interpolating flows in $(1+1)$ hydrodynamics }}\\

\vs{10}

{\large
Aritra Banerjee${}^{a,\,}$\footnote{\url{aritra.banerjee@pilani.bits-pilani.ac.in}}, Abir Ghosh${}^{b\,}$\footnote{\url{abirghosh.physics@gmail.com}}, \\ 

\vskip 0.3in

{ {Birla Institute of Technology and Science, Pilani Campus, Rajasthan 333031, India}$^{a}$ }\vskip .5mm

{ {Indian Institute of Science, Bangalore 560012, India}$^{b}$}\vskip .5mm

{\it ${}^{}$}\vskip.5mm}

\end{center}

\vskip 0.35in

\begin{abstract}
We revisit the general analytic solution space for relativistic $(1+1)$-dimensional hydrodynamics for a perfect fluid flowing along the longitudinal direction.  We work out the explicit one-parameter family of interpolating flows between boost-invariant and boost-non-invariant regimes, where a direct and simple dialing of the parameter at the level of solutions is possible. We also discuss the construction of generalised rapidity distribution of entropy for such interpolating flows at the level of potentials.
	\end{abstract}

	\newpage

\tableofcontents

\section{Introduction}
Relativistic hydrodynamics has been widely successful in describing collective phenomena in high energy QCD and early universe physics. Based on experimental data obtained from relativistic Heavy Ion Collisions (HICs) conducted both at the RHIC and LHC, we can confirm that a hot and dense nuclear matter is generated during the initial stages of these collisions, commonly referred to as Quark-Gluon Plasma (QGP). It is astonishing that many physical phenomena associated to QGP is well approximated by the almost perfect, strongly-coupled fluid picture of high energy particle processes in these regimes. The problem of solving hydrodynamics in $(1+1)$ dimensions has subsequently receieved particular attention since for very high energy particles, rotational invariance is approximately preserved along the transverse directions\footnote{However, rotational symmetry, while keeping the motion independent of azimuthal coordinates, can still allow radial expansions, making the actual system $(1+2)$ dimensional. }. Depending on the rapidity dependence of such a flow, one could classify the associated heavy-ion collision processes. 

In general, there are two broad classes of solutions deemed most important in this regime, namely the ones with and without manifest boost invariance. The boost invariant regime for a perfect fluid was developed by Bjorken \cite{PhysRevD.27.140} (see also \cite{Hwa:1974gn}) in his seminal work, which applies beautifully to the central rapidity region of high energy collisions of relativistic heavy ions. Due to the symmetry protected nature, one can easily determine the four-velocity profile for the beam in question. Although elegant in formulation, the Bjorken boost invariant predictions are hardly ever important for experiments. In such cases, still considering a highly idealised situation, Gaussian rapidity distributions are thought to dominate the data for the dense hot matter produced in high energy collisions, giving rise to what is know as the Landau flow \cite{Landau:1953gs}. This has been experimentally observed in high energy scattering events \cite{Steinberg:2004vy, Wong:2008ta, BRAHMS:2004dwr,PHENIX:2004vcz,PHOBOS:2004zne,STAR:2005gfr}\footnote{More extensive discussions on Landau flows appear in a number of works, including \cite{misc1,misc2,misc3,misc4,misc5,misc6,misc7}. This list is by no means exhaustive and the interested reader is requested to look at the references of these.}. However lucrative, exact analytic solutions of fluid equations in $(1+1)$ dimensions remain scarce, and general solutions spanning all parameter space is still hard to find even for benchmarking purposes. 

Although given by very different regimes of parameter space in the fluid theory, both Bjorken and Landau flow share something unique, in the sense they are obtained assuming the rapidity $y$ for the flow satisfies a harmonic equation for the lightcone kinematic variables, giving them the epithet ``Harmonic flows". Using this connection, a family of solutions interpolating between Bjorken and Landau regime (a generalised ``in-out ansatz") was put forward in  \cite{PhysRevC.76.054901}. In this work the authors found out a way to write a solution for the $(1+1)$ perfect fluid theory with arbitrary speed of sound, that is valid in boost non-invariant regimes, but the invariance can be restored in a particular limit. This structure is the main focus of the current manuscript. 

In this work, we revisit the interpolating solution between the Bjorken and Landau regime, taking extra care to provide a dialing parameter that seamlessly transform observables between the two. We explicitly derive the one parameter dependent ansatz for the general solution which works in both regimes. Further, we construct an exact interpolating entropy flow valid for both regimes by using the formalism of the Khalatnikov potential \cite{khal}. The Khalatnikov formalism allows one to replace the non-linear problem of hydrodynamic evolution with linear differential equations for a suitably defined scalar potential. This gives us analytic control over the solution space in terms of temperature and rapidity, which we then show can be used to define physical observables over a larger parameter space.  

The rest of the discussion is structured as follows: in section \eqref{sec:2} we discuss some background material on hydrodynamic equation in lightcone coordinates. In section \eqref{sec:3} we rework the construction of the interpolating ansatz from the solution of these equations, which depends on a parameter that appears as an integration constant. In section \eqref{sec:4}  we use this ansatz to find out the entropy flow as function of the freeze-out temperature. In section \eqref{sec:5} we further extend the parameter space of our ansatz to take care of other classes of harmonic flow solutions. We end with some discussions in \eqref{sec:6}.

\section{The set-up: Hydrodynamic equations}\label{sec:2}
Let us first review the basics of the formulation in a two dimensional $(t,z)$ plane. 
We start by setting up the basic hydrodynamic equations in the light cone variables given by $\z = t \pm z \,$,
a coordinate choice appropriate for particles moving with speeds comparable to $c$, which we will set in our convention to be fixed at 1 for the rest of the manuscript. 
The perfect fluid energy momentum tensor is given by,
\begin{equation}
T^{\mu\nu} = \left(\epsilon + p\right) u^\mu u^\nu + p \eta^{\mu\nu}\, , \label{eq:stress}
\end{equation}
where, $\epsilon$ is the energy density, $p$ is the pressure of the fluid and $u^\mu$ is the four-velocity with $u^\mu u_\mu = -1$. We also assume a linear equation of state connecting the two given by,
\begin{eqnarray}\label{eq:eq_of_state}
    \epsilon=g p
\end{eqnarray}
where $1/\sqrt{g}$ is speed of sound in the fluid, and not to be confused with the metric tensor. Finally, the hydrodynamic equation is nothing but the local conservation laws,
\begin{equation}
    \partial_\mu T^{\mu\nu} = 0. \label{eq:hydroeqn}
\end{equation}
With the light-cone coordinates the partial derivatives become,
    $\partial_{\pm}=\frac{\partial}{\partial \z}\equiv \frac{1}{2}\left(\frac{\partial}{\partial t}\pm\frac{\partial}{\partial z}\right)$
and the hydrodynamic equation takes the form,
\begin{equation}\label{eq:em_eq}
    \partial_{\pm} T^{01} + \frac{1}{2}\partial_+(T^{11} \pm T^{00}) - \frac{1}{2}\partial_-(T^{11} \mp T^{00})=0 .
\end{equation}
We also express the velocities in the light cone coordinate as $u^{\pm} \equiv u^0 \pm u^1 = e^{\pm y}$; where $y$ is the rapidity, defined as,
\begin{equation}\label{eq:rapidity}
    y \equiv \frac{1}{2}\ln\left(\frac{\epsilon+p_z}{\epsilon-p_z}\right) = \frac{1}{2}\ln\left(\frac{1+v}{1-v}\right),
\end{equation}
where $p_z$ is the momentum along $z$ direction (direction of the flow). Substituting the definition of velocity and \eqref{eq:eq_of_state} in \eqref{eq:em_eq} we get,
\begin{subequations}\label{eq:EOM}
    \begin{align}
        g \partial_+ \ln p &= - \frac{(g+1)^2}{2} \partial_+y - \frac{g^2-1}{2} e^{-2y} \partial_-y\\
        g \partial_- \ln p &=  \frac{(g+1)^2}{2} \partial_-y + \frac{g^2-1}{2} e^{2y} \partial_+y
    \end{align}
\end{subequations}
Notice that these two equations are enough to fix all physical variables associated to the state of the fluid. In general these sets of equations could be highly non-linear, but we may still be able to get some exact solutions.
Eliminating $p$ using the above equations; we get a consistency condition for $y$,
\begin{equation}\label{eq:condition}
    \partial_+\partial_- y=\frac{g^2\!-\!1   }{4 (1\!+\!g)    ^2}\left\{\partial_-\partial_-[e^{-2y}]-\partial_+\partial_+[e^{+2y}]\right\}
\end{equation}
which in general can be thought of as a wave equation for the rapidity with a source term.
This consistency condition will give us the differential equation for finding out a general fluid flow profile. These sets of equations \eqref{eq:EOM}-\eqref{eq:condition} will be our main focus in this manuscript.

\section{Interpolating flows: Boost and beyond}\label{sec:3}
Before jumping to the general interpolating ansatz we will look at the Bjorken flow which would help us build some intuition for the ansatz. Bjorken flow \cite{PhysRevD.27.140} is a statement about the velocity profile of the fluid flow. Bjorken flow fluid profile is given by,
\begin{equation}\label{eq:bjorken_profile}
    v(z,t)=\frac{z}{t}
\end{equation}
where, $v(z,t)$ is the velocity of the fluid at the point $(z,t)$. We urge the readers to be mindful, that \eqref{eq:bjorken_profile} is the \textit{definition} of Bjorken flow and is slightly different from \eqref{eq:rapidity}. The later is true for any relativistic particle with velocity $v$. Substituting the above fluid profile in \eqref{eq:rapidity} we get,
\begin{equation}
     y = \frac{1}{2}\ln\left(\frac{\epsilon+p_z}{\epsilon-p_z}\right) = \frac{1}{2}\ln\left(\frac{t+z}{t-z}\right).
\end{equation}
People often re-write the above equation as $y=\eta$ and use this as the definition of Bjorken flow. Here, $\eta$ is the ``space-time rapidity" and defined as,
\begin{equation}
    \eta= \frac{1}{2}\ln\left(\frac{t+z}{t-z}\right).
\end{equation}
In the light-cone velocity variables the equation looks like,
\begin{equation}\label{eq:bjorken_ansatz}
    2y=\ln{u_{+}}-\ln{u_{-}}=\ln{z_{+}}-\ln{z_{-}}
\end{equation}
which signifies the equivalence of rapidity and space-time rapidity of the fluid flow.  This simple ansatz makes sure that the system is manifestly boost invariant, i.e. put into \eqref{eq:EOM} the pressure of the flow and other thermodynamic quantities do not depend on the rapidity. 
The form of the above equation naturally motivates the generalised ansatz given by the following equation,
\begin{equation}\label{eq:gen_ansatz}
    2y=\ln{u_{+}}-\ln{u_{-}}=\ln{f_{+}(z_{+})}-\ln{f_{-}(z_{-})}
\end{equation}
were $f_{\pm}$ are arbitrary functions of the light-cone coordinates. Note that this is essentially a more general statement about the fluid flow profile. With $f_\pm = z_\pm$ we get back the Bjorken flow. It should be immediately clear that a generic $f_\pm$ will break the boost invariance associated to Bjorken ansatz. This non boost-invariant regime is sometimes called Landau flow \cite{Landau:1953gs} as discussed in the introduction.

What remains to be done is to constrain forms of $f_\pm$ via the hydrodynamic equations. Substituting this ansatz in the consistency equation \eqref{eq:condition} we get for the functions $f_{\pm}$,
\begin{equation}\label{eq:dif_f}
        f_-\partial_-\partial_-(f_-)=f_+\partial_+\partial_+(f_+)= A^2/2 
\end{equation}
Thus both $f_+$ and $f_-$ satisfy the identical equation,\footnote{We drop the $\pm$ subscript of $f$ whenever the equations are separately true for both. Primes would now on denote derivatives w.r.t $z_\pm$ as per the case. }
\begin{equation}\label{diffeq}
    f f''=A^2/2,
\end{equation}
and $A$ is a constant parameter. 
This in turn can be written as ,
\begin{subequations}\label{eq:C_factor}
    \begin{align}
        &[(f')^2]'= A^2 [\ln f]' \\
        \Rightarrow \qquad & f'  = \sqrt{A^2 \ln f + C} \label{eq:f}
    \end{align}
\end{subequations}
One can easily see that in the limit $A\rightarrow0$ the function $f$ becomes linear in $z$, which is evident from the differential equation \eqref{diffeq} as well. This is our first encounter with an interpolating solution, that reduces down to Bjorken flow. 

Now we arrive at a crucial juncture in our computation. Since we want the solution for $f$ to go to eq\eqref{eq:bjorken_ansatz} explicitly as we dial $A$ to zero, we will set $C=1$. Note, that this is where we differ with the authors as discussed in the original paper \cite{PhysRevC.76.054901}, as for their case, a smooth interpolation by dialing $A$ only seems not possible.   
\footnote{The authors of \cite{PhysRevC.76.054901} set $C=A^2\log{\left(H\right)}$,
where $H$ is some arbitrary constant and get,
\begin{equation}
    f'  = A\sqrt{\ln{\left(\frac{f}{H}\right)}}
\end{equation}
One can immediately see that we cannot recover the Bjorken flow at the level of solutions because as $A\rightarrow0$ the integration constant $C$ also vanishes and instead of eq\eqref{eq:bjorken_ansatz} we end up with a constant $f$.}
So we finally have a new implicit form for the solution,

\begin{equation}\label{eq:int_f}
    z = \int \frac{df}{\sqrt{A^2 \ln f + 1}}
\end{equation}
This integral can be solved by brute force which gives us a complicated function which we will present later. One can also expand the integrand in orders of $A$ and then integrate to get results perturbatively away from the Bjorken regime. Doing that we get,
\begin{equation}\label{eq:expansion}
    z=f-\frac{1}{2} f (\ln (f)-1)\; A^2 +\frac{3}{8}  f \left(\ln ^2(f)-2 \ln (f)+2\right)\;A^4+O\left(A^6\right),
\end{equation}
which re-confirms that in the $A\rightarrow0$ limit we do get back the linear term \footnote{Interested readers can check that the above is the expansion one gets if one expands the exact solution.}. Another point to be made here is about the dimensionality of $A$. Since rapidity $y$ is a dimensionless parameter, eq\eqref{eq:gen_ansatz} tells us that we have freedom to choose the dimensions of the function $f$.\footnote{If we are being stringent then in \eqref{eq:int_f} we need to add another term to make it dimensionally correct. That is, $z = \int (\sqrt{A^2 \ln (f/f_0) + 1})^{-1/2}df$. Or one can also just work with a dimensionless $f$ to begin with.} Consequently the dimensions of $A$ from \eqref{eq:dif_f} is: $\left[A\right]=\left[f\right]\left[L^{-1}\right]$. Since for Bjorken flow the ansatz is simply $f_\pm=z_\pm$, a natural choice is to set $f$ to be of the dimensions of length which makes our interpolation parameter to be dimensionless; which is also intuitive to some extent since the $A$ gives us an effective dial. All in all, we now have a one-parameter family of hydrodynamic solutions that should interpolate, according to the value of $A$, between boost-invariant and non-invariant regimes.

Now that we have an idea about explicit form of $f$ we can go ahead with calculating the thermodynamic properties of the fluid \footnote{One can also get the functional form by just integrating \eqref{eq:int_f}. The function is presented later in \eqref{eq:new_eq}.}. Substituting the ansatz \eqref{eq:gen_ansatz} in \eqref{eq:EOM} we get,
\begin{align}
    g\partial_+[\ln p]&=-\frac{ (1\!+\!g)    ^2}4\frac{f_+'}{f_+}+\frac{g^2\!-\!1   }4\frac{f_-'}{f_+}\\
    g\partial_-[\ln p]&=-\frac{ (1\!+\!g)    ^2}4\frac{f_-'}{f_-} +\frac{g^2\!-\!1   }4\frac{f_+'}{f_-}
\end{align}
From this we get by integrating,
\begin{subequations}\label{24}
    \begin{align}
            g\ln p&=-\frac{ (1\!+\!g)    ^2}4\ln f_+ +\frac{g^2\!-\!1   }4f_-'\int \frac{dz_+}{f_+}+ \Delta_-(z_-) \\
            g\ln p&=-\frac{ (1\!+\!g)    ^2}4\ln f_- +\frac{g^2\!-\!1   }4f_+'\int \frac{dz_-}{f_-}+ \Delta_+(z_+),
    \end{align}
\end{subequations}
with $\Delta_{\pm}(z_\pm)$ being undetermined functions. 
The integrals on the R.H.S. can be evaluated using the explicit form of the function in \eqref{eq:f}. Indeed:
\begin{equation}
    \int \frac{dz}{f}=\int \frac{df}{f}\frac{1}{f'}=\int \frac{d(\ln(f))}{\sqrt{A^2 \ln f + 1}}=\frac{2 \sqrt{A^2 \ln f + 1}}{A^2} + K=\frac{2( \sqrt{A^2 \ln f + 1}-1)}{A^2}
\end{equation}
for the integral to be finite\footnote{This can be easily seen by expanding the expression in orders of $A$.} as $A \rightarrow 0$ we set the value of integration constant $K = -2/A^2$. Substituting this we get,
\begin{subequations}
    \begin{align}
            g\ln p&=-\frac{ (1\!+\!g)    ^2}4\ln f_+ +\frac{g^2\!-\!1   }{2}\frac{( \sqrt{A^2 \ln f_- + 1}\sqrt{A^2 \ln f_+ + 1}-\sqrt{A^2 \ln f_- + 1})}{A^2} + \Delta_-(z_-) \\
            g\ln p&=-\frac{ (1\!+\!g)    ^2}4\ln f_- +\frac{g^2\!-\!1   }{2}\frac{(\sqrt{A^2 \ln f_+ + 1} \sqrt{A^2 \ln f_- + 1}-\sqrt{A^2 \ln f_+ + 1})}{A^2} + \Delta_+(z_+)
    \end{align}
\end{subequations}
Comparing these to the expressions in \eqref{24} we can fix $\Delta_{\pm}(z_\pm)$:
\begin{equation}
    \Delta_\pm(z_\pm) =-\frac{ (1\!+\!g)    ^2}4\ln f_\pm + \frac{g^2\!-\!1   }{2}\frac{\sqrt{A^2 \ln f_\pm + 1}}{A^2} - \frac{g^2\!-\!1   }{2}\frac{1}{A^2},
\end{equation}
which also has the explicit information of $A$.
Finally collecting all the terms gives us,
\begin{equation}
     g\ln p=-\frac{ (1\!+\!g)    ^2}4\ln (f_+ f_-) +\frac{g^2\!-\!1   }{2}\frac{( \sqrt{A^2 \ln f_- + 1}\sqrt{A^2 \ln f_+ + 1}-1)}{A^2}
\end{equation}
Using all of this, we can solve for the pressure and the rapidity in terms of the lightcone variables,
\begin{subequations}\label{eq:full_p_and_y}
    \begin{align}
           p(z_+,z_-) &= p_0 \exp{\left\{-\frac{ (1\!+\!g)    ^2}{4g}\ln (f_+ f_-) +\frac{g^2\!-\!1   }{2g}\frac{( \sqrt{A^2 \ln f_- + 1}\sqrt{A^2 \ln f_+ + 1}-1)}{A^2}\right\}} \\
           y(z_+,z_-) &= \frac{1}{2}\ln{\left(\frac{f_+}{f_-}\right)}
    \end{align}
\end{subequations}
The general solution of course depends on the rapidity variables. However, the solution for pressure in the limit $A \rightarrow 0$ goes to\footnote{We have explicitly added the $\tau_0$ term for dimensional correctness in the expression for pressure.},
\begin{equation}
    p(z_+,z_-)= p_0\;(z_+z_-)^{-(g+1)/2g} = p_0~\left(\frac{\tau}{\tau_0}\right)^{-(g+1)/g}
\end{equation}
with $\tau$ as proper time, this is nothing but the solution for the Bjorken ansatz, as the boost invariance of physical quantities is evident.

Now note that the generic form of functions $f_\pm$ can be integrated to get the following form,
\begin{equation} \label{eq:new_eq}
    f_{\pm} = \exp\left\{\left(\text{erfi}^{-1}\left(\frac{A\; z_{\pm}}{\sqrt{\pi}}e^{\frac{1}{A^2}}  \right) \right)^2-\frac{1}{A^2}\right\}
\end{equation}
Here erfi$^{-1}$ signifies an inverse imaginary error function\footnote{This is defined as the negative imaginary part of the regular real error function evaluated at an imaginary argument. Despite the name the function is real when the argument is real.}. Even though not obvious at first, we can see these functions $f_\pm$ are staggeringly difficult to expand near $A=0$, but the behavior becomes more understandable from the expansion \eqref{eq:expansion}: the presence of $\ln(f)$ terms at every order makes inverting the series very difficult as the equations become transcendental, but at the leading order we just have $f=z$ near $A=0$. This fact can be checked by direct plotting of the generic solution as well. Note that on the lightcones, i.e. $z_\pm = 0$, this function is just $e^{-\frac{1}{A^2}}$. This means the pressure becomes a pure constant on the lightcones and the flow becomes stationary.

\begin{figure}[H]
    \centering
    \includegraphics[width=12cm]{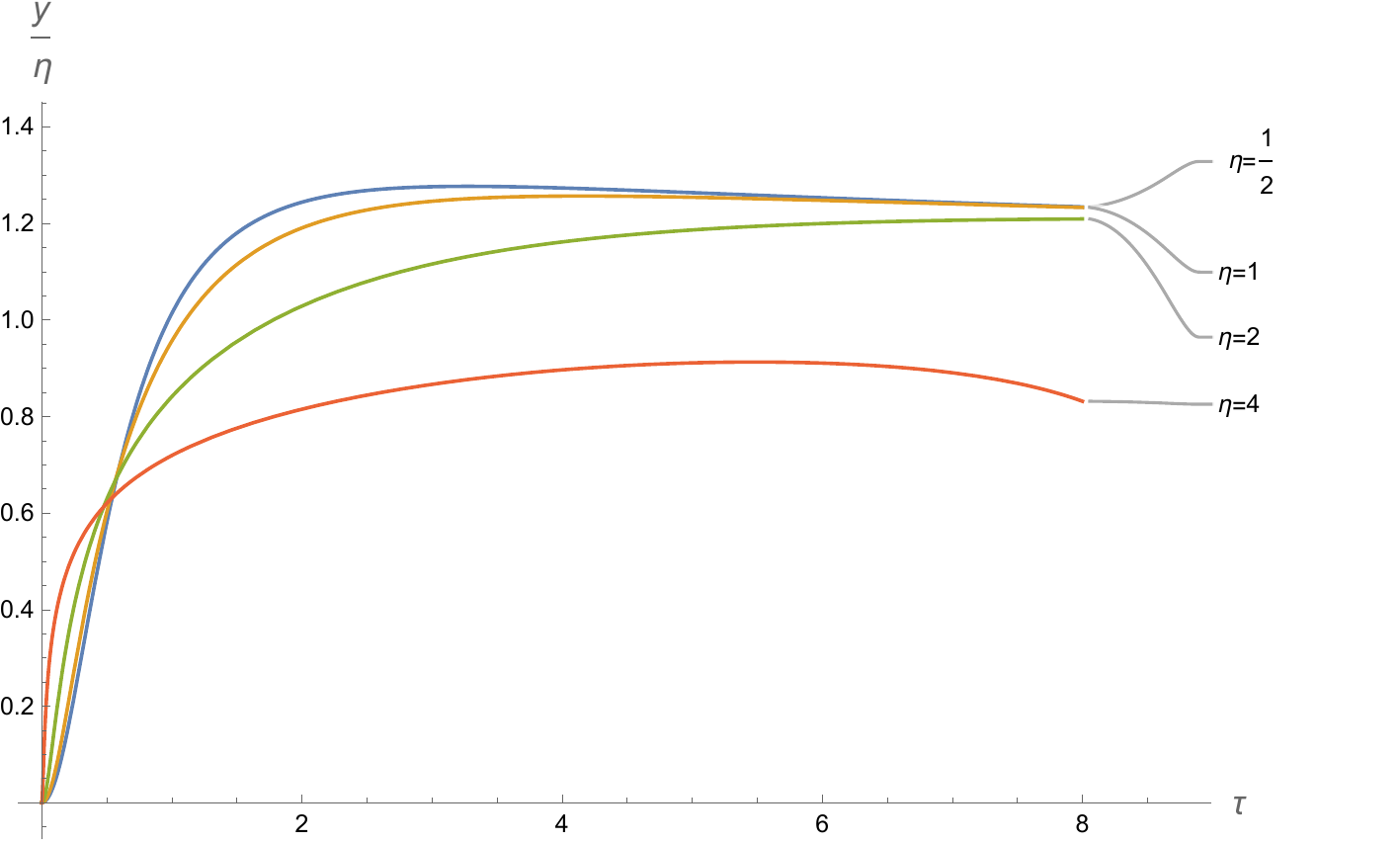}
    \caption{The ratio of $y/\eta$ has been plotted against the proper time $\tau$ for a constant $A$ and varying rapidity $\eta$. One can see that for large proper time all graphs show boost invariance as $y/\eta$ asymptotically goes to one.}
    \label{fig:Figure_2}
\end{figure}

\begin{figure}[h]
    \centering
    \includegraphics[width=12cm]{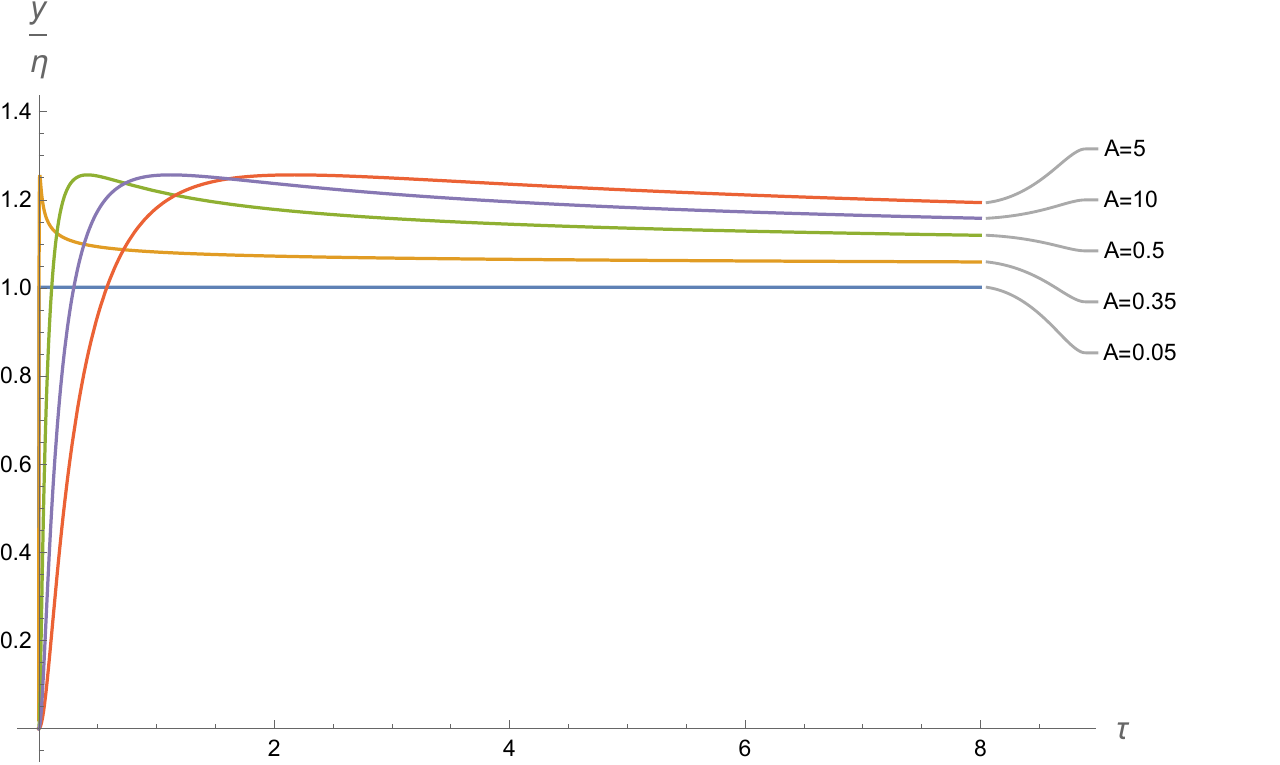}
    \caption{The ratio of $y/\eta$ has been plotted against the proper time $\tau$ for a constant $\eta$ and varying interpolation parameter $A$. Some we see that for smaller and smaller $A$ values the graph becomes flatter around 1 which is what we expect since we move towards the Bjorken flow limit. }
    \label{fig:Figure_3}
\end{figure}

Finally we present the solution in a form often used in literature,
\begin{subequations}\label{eq:therm_var}
    \begin{align}
           p(z_+,z_-) &= p_0 \exp{\left\{\frac{1}{A^2}\left[-\frac{ (1\!+\!g)^2}{4g}(l_-^2+l_+^2) +\frac{g^2\!-\!1}{2g}l_-l_+ + \frac{g+1}{g}\right]\right\}} \\
           y(z_+,z_-) &= \frac{1}{2 A^2}\left(l_+^2-l_-^2\right)
    \end{align}
\end{subequations}
where, $l_\pm=\sqrt{A^2 \ln{f_\pm}+1}$. Here, we have tried to keep the form of the expressions similar to the original work \cite{PhysRevC.76.054901}; but, note that the definition of $l_\pm$ is different here, and so is the extra $\frac{g+1}{g}$ in the expression of pressure. But more importantly, these expressions can be seamlessly interpolated to the boost invariant regime. To round out this discussion, we show how the general solution based on \eqref{eq:new_eq} is different from the boost invariant regime, via plotting the value of $y/\eta$ for varying proper time and varying $A$ in Fig. \ref{fig:Figure_2} and \ref{fig:Figure_3}.

With the explicit form of the function $f_{\pm}$ in \eqref{eq:new_eq} we can now also write down the explicit velocity profile of the interpolating flow; but, since the form of $f_{\pm}$ isn't pretty, consequently the functional form of the velocity profile is not very enlightening and we refrain from mentioning it here.

\section{Entropy flow}\label{sec:4}
Now that we have the pressure in terms of the lightcone variables and $f_\pm$, one can calculate other thermodynamic quantities using the standard relations,\footnote{We have assumed vanishing chemical potential.}
\begin{equation}\label{eq:therm_rel}
    p+\epsilon=Ts\;;\;d\epsilon=T ds\;;\;dp=sdT
\end{equation}
where, $p,\epsilon \text{ and }s$ are pressure, energy density and entropy density, respectively. The thermodynamic relations here are closed using the equation of state \eqref{eq:eq_of_state}; which essentially means that the speed of sound is held constant in the analysis,
\begin{equation}
    \frac{dp}{d\epsilon}=\frac{s}{T}\frac{dT}{ds}=c_s^2\equiv \text{const.}
\end{equation}
Throughout the calculation the quantities $p,\epsilon \text{ and }s$ are considered in the rest frame of the fluid. With above equations one gets the following thermodynamic relations between the variables. The constant speed of sound is as mentioned before:
$c_s^2\equiv\frac{p}{\epsilon}=\frac{1}{g}$.
We also define the quantity $\theta$, related to the temperature profile, as:
\begin{equation}
    \theta=\log{\left(\frac{T_0}{T}\right)}=\frac{1}{g+1}\log{\left(\frac{p_0}{p}\right)}
\end{equation}
where $p_0\equiv\frac{\epsilon_0}{g}T_0^{g+1}$. In this language the energy density is,
\begin{equation}
\epsilon=gp=\epsilon_0\left(\frac{T}{T_0}\right)^{g+1}=\epsilon_0e^{-(g+1)\theta}
\end{equation}
and for entropy we get,
\begin{equation}
s=s_0\left(\frac{T}{T_0}\right)^{g}=s_0e^{-\theta}.
\end{equation}
One is generally interested in the distribution of entropy per unit rapidity $ds/dy$ as it is a relatively easily measurable quantity and further, it may be related to the multiplicity distribution of particles. Since we have already found an one-parameter interpolating flow in the last section, it would be only natural to look for a general interpolating formula for entropy distribution as well. 

This ``hydrodynamic observable" depends on the presumed $(1+1)$ dimensional hypersurface through which the flow is considered. One can consider all sorts of surfaces such as a fixed proper time surface or fixed time surface. A particularly interesting hypersurface is the one corresponding to a fixed temperature. It allows us to follow the evolution of the fluid from a high temperature inital state towards a final low temperature state which is often associated to the \textit{freeze out} temperature. It essentially is the temperature/surface where the system starts to lose its fluid nature and starts behaving more like particles. 

\begin{figure}[]
    \centering
    \includegraphics[width=0.92\linewidth]{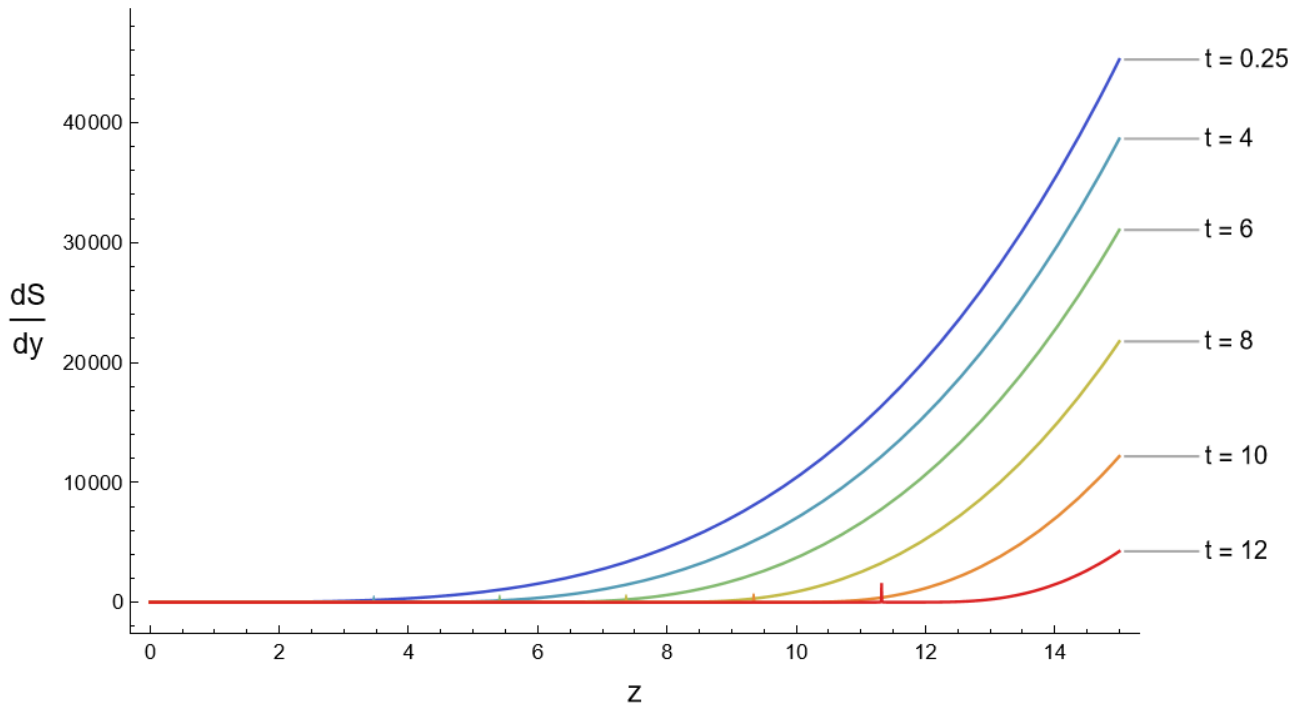}
    \caption[foot]{Plot of the entropy flow distribution for a non-boost invariant flow in \eqref{eq:entropy_flow_dist} at different time instances, keeping other parameters fixed.
    The redder curves are the plots for later time. We see that the distribution flattens as the fluid flow evolves.}
    \label{fig:entropy_flow}
\end{figure}

\begin{figure}[]
    \centering
    \includegraphics[width=0.9\linewidth]{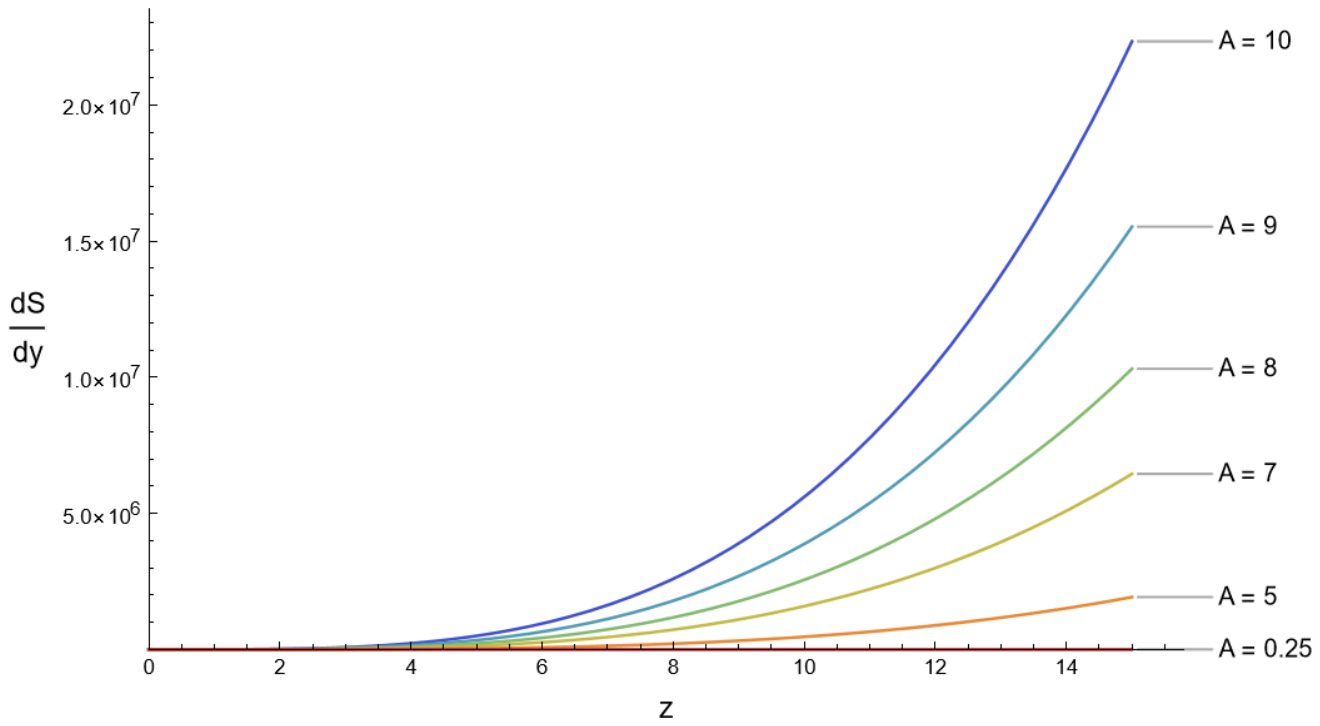}
    \caption[note]{Plot of the entropy flow distribution in \eqref{eq:entropy_flow_dist} for a non-boost invariant flow for different value of $A$, keeping other parameters fixed.
    The redder curves are the plots for smaller $A$ value. We see that the distribution flattens for small values of $A$, as expected.}
    \label{fig:diff_val_A}
\end{figure}

We go about calculating the entropy distribution using the method of \textit{Khalatnikov potentials} \cite{khal} \footnote{See the english translation of the original paper \cite{Belenkij:1955pgn}.} as the calculations are much simpler in this case, and further allows to devise exact solutions for entropy flow in $(1+1)$ hydrodynamics. We start from \eqref{eq:em_eq} and using the thermodynamic relations in \eqref{eq:therm_rel}, we can write,
\begin{equation}
    \partial_+\left(e^{-\theta+y}\right)=\partial_-\left(e^{-\theta-y}\right) \equiv \partial_+\partial_-\Phi\left(\zp,\zm\right)
\end{equation}
The first equality above can to be seen as a curl of a the quantity $\Vec{Q}=(e^{-\theta-y},e^{-\theta+y}) = (u^{-}T,u^{+}T)$ being put equal to zero. Therefore, $\Vec{Q}$ can be written as a divergence of a scalar potential $\Phi$ as above. The consistency relation then dictates:
\begin{equation}
    \partial_\pm\Phi(\zp,\zm)\equiv u^{\mp}T=T_0 e^{-\theta\mp y}.
\end{equation}
This scalar $\Phi$ can be related to another potential via a Legendre transformation, called the Khalatnikov potential. However, as argued in \cite{PhysRevC.78.064909, Peschanski:2010cs}, one can derive all physical quantities from $\Phi$ itself.
Substituting $p$ from our interpolating solution \eqref{eq:therm_var}  we get in the $(\theta,y)$ variables,
\begin{subequations}\label{eq:ty}
    \begin{align}
        \theta &= \frac{1}{A^2}\left[\frac{ (1\!+\!g)}{4g}(l_-^2+l_+^2) -\frac{g\!-\!1}{2g}l_-l_+ - \frac{1}{g}\right] \\
        y &= \frac{1}{2 A^2}\left(l_+^2-l_-^2\right)
    \end{align}
\end{subequations}
To calculate the scalar potential $\Phi$ we write,
\begin{subequations}\label{eq:phi_diff}
    \begin{align}
        \frac{\partial\Phi}{\partial l_\pm}&=\frac{\partial z_\pm}{\partial l_\pm}\frac{\partial\Phi}{\partial z_\pm}=\frac{2T_0}{A^2}\exp{\left(\frac{l_\pm^2-1}{A^2}-\theta\mp y\right)}\\
        &=\frac{2 T_0}{A^2}\exp{\left( \frac{g-1}{4 A^2 g} \left(l_+ + l_-\right)^2 -\frac{g-1}{A^2 g} \right)}
    \end{align}
\end{subequations}
where we have used \eqref{eq:int_f} and \eqref{eq:therm_var}. This gives us a differential equation which can be easily solved to get an interpolating version of the potential, again in terms of imaginary error functions:
\begin{equation}
    \Phi= 2\sqrt{\pi} e^{ -\frac{(g-1)}{A^2 g}} \frac{T_0}{A}  \sqrt{\frac{g}{g-1}}~\mathrm{erfi}\left[\frac{1}{2A}\sqrt{\frac{g-1}{g}}(l_+ + l_-)\right]
\end{equation}
$l_+ + l_-$ in the above equation can be calculated from \eqref{eq:ty} to be,
\begin{equation}
    l_+ + l_-=\sqrt{2g}\left(A^2\theta+\frac{1}{g}+\sqrt{\left(A^2\theta+\frac{1}{g}\right)^2-\frac{\left(A^2y\right)^2}{g}}\right)^{1/2}
\end{equation}
Finally, as discussed in \cite{PhysRevC.78.064909}, one can substitute the potential into the following relation to get entropy per unit rapidity as a function of the temperature,
\begin{align}
        \frac{\mathrm{d}S}{\mathrm{d}y}\left(\theta,y\right)&=\frac{s_0}{2gT_0}e^{-(g-1)\theta}\partial_\theta\Phi(\theta,y)=\frac{s_0}{2gT_0}e^{-(g-1)\theta} \left(\frac{\partial l_+}{\partial \theta}\frac{\partial\Phi}{\partial l_+}+\frac{\partial l_-}{\partial \theta}\frac{\partial\Phi}{\partial l_-}\right)\nonumber\\
        &=\frac{s_0}{\sqrt{2g}}\exp{\left[\frac{g-1}{2A^2}\left(\left(\frac{1}{g}+A^2\theta\right)+\left(\left(\frac{1}{g}+A^2\theta\right)^2-\frac{A^4y^2}{g}\right)^{1/2}-4\right)\right]}\nonumber\\
        &\hspace{5cm}\left(1+\left(\frac{1}{g}+A^2\theta\right)\left(\left(\frac{1}{g}+A^2\theta\right)^2-\frac{A^4y^2}{g}\right)^{-1/2}\right)\label{eq:entropy_flow_dist}
\end{align}
Note that this is in general a gaussian Landau-like distribution in rapidity. The evolution of this entropy flow for different times and different $A$'s have been shown in Fig. \ref{fig:entropy_flow} and Fig. \ref{fig:diff_val_A} respectively.

With this we have derived a one parameter expression of entropy flow over a constant temperature surface that interpolates between Bjorken and Landau regimes. Taking an $A\rightarrow0$ limit of this expression is not trivial. But, an easier way is to take the limit earlier in the derivation (say \eqref{eq:ty}); with this one gets the following expression for the potential at the Bjorken regime,
\begin{equation}
    \Phi=T_0\tau_0\frac{2g}{(g-1)}\exp{\left((g-1)\theta\right)}
\end{equation}
And the entropy flow as expected is constant in this limit,
\begin{equation}
    \frac{\mathrm{d}S}{\mathrm{d}y}\left(\theta,y\right)=s_0\tau_0
\end{equation}
This serves as a check for our calculation for the general potential. 

\section{General harmonic flow solutions}\label{sec:5}
As discussed in preceding sections, the solution \eqref{eq:full_p_and_y} and \eqref{eq:new_eq} are one parameter solutions for the equation \eqref{eq:condition}. But we can actually do better. The anstaz \eqref{eq:gen_ansatz} solves the equation \eqref{eq:condition} but a homogeneous version thereof,
\begin{equation}
    \partial_+\partial_-y=(\partial_t^2-\partial_z^2)y=0.
\end{equation}
Such family of flows, as alluded to in the introduction, are called ``Harmonic flows" as physical rapidity is a harmonic function of the lightcone kinematic variables. Note that in our setup both $l_\pm(z_\pm)$ are harmonic functions and so is $y$ by consequence. If we keep the integration constant $C$ in \eqref{eq:C_factor}  instead of fixing it, we could get a bigger family of solutions parameterized by both $A$ and $C$ values. With the constant $C$ explicitly carried along, eq \eqref{eq:new_eq} and \eqref{eq:therm_var} become,
\begin{equation} 
    f_{\pm}= \exp\left\{\left(\text{erfi}^{-1}\left(\frac{A\; z_{\pm}}{\sqrt{\pi}}e^{\frac{C}{A^2}}  \right) \right)^2-\frac{C}{A^2}\right\},
\end{equation}
and similarly for other quantities:
\begin{subequations}
    \begin{align}
           p(z_+,z_-) &= p_0 \exp{\left\{\frac{1}{A^2}\left[-\frac{ (1\!+\!g)^2}{4g}(l_-^2+l_+^2) +\frac{g^2\!-\!1}{2g}l_-l_+ + C~\frac{g+1}{g}\right]\right\}} \\
           y(z_+,z_-) &= \frac{1}{2 A^2}\left(l_+^2-l_-^2\right)
    \end{align}
\end{subequations}
where we use new variables $l_\pm=\sqrt{A^2 \ln{f_\pm}+C}$. Note immediately that if we could put $C=0$, the $\frac{g+1}{g}$ term in the pressure goes away. 

One can also calculate the entropy flow with these larger family of harmonic flow solutions. Using the above equations we get,
\begin{equation} \label{50}
    \frac{\partial\Phi}{\partial l_+}=\frac{\partial\Phi}{\partial l_-}=\frac{2T_0}{A^2}\exp{\left(\frac{g-1}{4gA^2}\left((l_++l_-)^2-4C\right)\right)}
\end{equation}
Finally, one can use this to get entropy per unit rapidity as a function of the temperature,
\begin{align}
        \frac{\mathrm{d}S}{\mathrm{d}y}\left(\theta,y\right)&=\frac{s_0}{2gT_0}e^{-(g-1)\theta}\partial_\theta\Phi(\theta,y)=\frac{s_0}{2gT_0}e^{-(g-1)\theta} \left(\frac{\partial l_+}{\partial \theta}\frac{\partial\Phi}{\partial l_+}+\frac{\partial l_-}{\partial \theta}\frac{\partial\Phi}{\partial l_-}\right)\nonumber\\
        &=\frac{s_0}{\sqrt{2g}}\exp{\left[\frac{g-1}{2A^2}\left(\left(\frac{C}{g}-A^2\theta\right)+\left(\left(\frac{C}{g}+A^2\theta\right)^2-\frac{A^4y^2}{g}\right)^{1/2}-4C\right)\right]}\nonumber\\
        &\hspace{5cm}\left(1+\left(\frac{C}{g}+A^2\theta\right)\left(\left(\frac{C}{g}+A^2\theta\right)^2-\frac{A^4y^2}{g}\right)^{-1/2}\right)
\end{align}

One can now reproduce the solutions presented in \cite{PhysRevC.76.054901} and \cite{PhysRevC.78.064909} by setting $A=1$ and $C=0$.\footnote{We emphasize this once again that with $C=0$ one cannot go to the Bjorken regime in any limit, not at least just by dialing parameters.}
\begin{equation}
    \frac{\mathrm{d}S}{\mathrm{d}y}\left(\theta,y\right)=\frac{s_0}{\sqrt{2g}}\exp{\left[\frac{g-1}{2}\left(-\theta+\left(\theta^2-\frac{y^2}{g}\right)^{1/2}\right)\right]}\left(1+\theta\left(\theta^2-\frac{y^2}{g}\right)^{-1/2}\right)
\end{equation}
In fact, in this limit, one notes that:
\begin{equation}
    l_+ + l_-=\sqrt{2g}\left(\theta+\sqrt{\theta^2-y^2/g}\right)^{1/2},
\end{equation}
which leads to the so-called singular behaviour at $y \to \pm c_s\theta$. 
We must note here that this is a very special instance in the solution space. In general any flow derived from \eqref{50} would be a Harmonic flow solution.

\section{Discussions and conclusions}\label{sec:6}
In this work, we discussed a refinement of the generalised fluid ansatz that unifies the Landau and Bjorken approaches to studying relativistic hydrodynamics. The full equations of motion can be solved for longitudinal flows effectively in $(1+1)$ dimensions, including free parameters which act as dials between different regimes of physical flows. We found out explicit expressions with arbitrary speed of sound for the physical observables in this case, and also worked out the entropy distribution in rapidity evaluated at different freeze-out temperatures. The harmonic property of the rapidity in hydrodynamic regime gives a large parameter space for viable solutions in this case.

This  note was merely to clarify the construction of the interpolating ansatz between boost invariant and non-invariant regimes. There are various questions and intricacies that remain in the discussion of exact solutions in $(1+1)$ hydrodynamics. An immediate question, for example, is the nature of the free parameter $A$. The equations \eqref{eq:bjorken_profile}, \eqref{diffeq}, and \eqref{eq:full_p_and_y} do not directly associate $A$ with a specific physical quantity. However, the plots in figures \eqref{fig:Figure_3}, and \eqref{fig:diff_val_A} where physical quantities vary with $A$ provide valuable insight into its physical interpretation. In principle this parameter measures the distance between the space-time rapidity and the kinematical rapidity, which coincide in the case of boost invariance.  As $A$ increases, the flow increasingly deviates from the Bjorken regime, indicating a breakdown of boost invariance. This could also be thought in terms of the rapidity width of the beam in question, since this quantity diverges at the Bjorken limit, which is uniform in rapidity. The more we add finite width corrections, the more we break boost invariance. In this sense $A$ could be thought of as a quantity related (perhaps invrsely) to rapidity widths as well, however one needs to investigate this in a better way.

A very interesting recent geometric perspective into the problem of Bjorken physics comes from Carrollian symmetries \cite{Bagchi:2023ysc}, which are basically speed of light going to zero limit of the relativistic cousin thereof \cite{SG, LBLL}. The highly ultra-relativistic nature of heavy-ion flows can be approximated as an effective $c\to 0$ expansion of relativistic hydrodynamics, where Bjorken flow has been shown to emerge at the leading order. In their derivation, the authors implement specific gauge choices for their frame fields and connections, that could be speculated to play a role analogous to the interpolation introduced by our parameter $A$. This connection suggests that $A$ may encode an effective velocity parameter, related to the geometric structures, thereby offering a richer interpretation beyond simple interpolation. This effective emergence of boost invariant flows from Carroll manifolds have also been extended to $3+1$ dimensional Gubser flows in \cite{Bagchi:2023rwd,Kolekar:2024cfg}. However, it is beyond the scope of the current work to connect this two disparate situations, both of which lead to Bjorken scalings in the limit. 

Outside of mathematical intrigues, non-boost invariant exact solutions are very important to construct. As we saw in this work, for the non-boost invariant regime, a generic ansatz for the $f_\pm$ could be solved for, however not all corners of the parameter space has been explored in the literature. The full solution works out as an imaginary error function, and one could find out other physically interesting regimes for the parameter space of the solution, for example the ones found in \cite{Csorgo:2018pxh}. Further, one may be able to connect to the discussion in \cite{Peschanski:2010cs}, where the authors found an infinite-dimensional linear basis of exact solutions that potentially classifies all families of solutions in $(1+1)$ dimensions. We hope to come back to some of these questions in future work.

\section*{Acknowledgements}
The authors would like to thank Arjun Bagchi, Arpan Das and Kedar Kolekar for useful related discussions. ABan is supported in part by an OPERA grant and a seed grant NFSG/PIL/2023/P3816 from BITS Pilani, and further an early career research grant ANRF/ECRG/2024/002604/PMS from ANRF India. He also acknowledges financial support from the Asia Pacific Center for Theoretical Physics (APCTP) via an Associate Fellowship. AG would like to thank BITS Pilani, Pilani Campus for kind hospitality during which a part of this work was done.

\bibliographystyle{utphysmodb}
\bibliography{ref.bib}

\end{document}